\documentclass[journal,comsoc]{IEEEtran}
\usepackage[pdftex]{graphicx}
\usepackage{epstopdf}
\usepackage[T1]{fontenc}
\usepackage[cmex10]{amsmath}
\usepackage{ifthen}
\usepackage{commath}
\usepackage[cmintegrals]{newtxmath}
\usepackage{esint}
\usepackage{cite} 
\usepackage{hyperref}
\usepackage{multirow}
\usepackage{xcolor}
\usepackage[font=small,labelfont=bf]{caption}
\newtheorem{remark}{Remark}
\newcommand{\dimv}[1]{{#1}^{\star}}    			
\newcommand{\kf}{k_f} 
\newcommand{\kb}{k_b}
\newcommand{\kd}{k_d}
 
\newcommand{\pr}{\text{Pr}}
\newcommand{\w}{\text{W}} 
\newcommand{\Pe}{\mathrm{P_e}}

\newcommand{\erfc}{\mathrm{erfc}\,} 

\newcommand{\pac}[1]{P_{AC}({#1}|r_0)}

\newcommand{\seq}{\mathbf{b}}
\newcommand{\rseq}{\mathbf{r}}
\newcommand{\bit}{b}
\newcommand{\diffd}[1]{D_{#1}} 
\newcommand{\TX}{\text{TX}}
\newcommand{\RX}{\text{RX}}

\begin{document}

\title{Diffusive Mobile Molecular Communications Over Time-Variant Channels}
\author{Arman~Ahmadzadeh,~\IEEEmembership{Student Member,~IEEE,}
        Vahid~Jamali,~\IEEEmembership{Student Member,~IEEE,} 
         Adam~Noel,~\IEEEmembership{Member,~IEEE,} 
        and~Robert~Schober,~\IEEEmembership{Fellow,~IEEE}\vspace{-6 mm}}
\maketitle 
\begin{abstract} 
This letter introduces a formalism for modeling time-variant channels for diffusive molecular communication systems. In particular, we consider a fluid environment where one transmitter nano-machine and one receiver nano-machine are subjected to Brownian motion in addition to the diffusive motion of the information molecules used for communication. Due to the stochastic movements of the transmitter and receiver nano-machines, the statistics of the channel impulse response change over time. We show that the time-variant behaviour of the channel can be accurately captured by appropriately modifying the diffusion coefficient of the information molecules. Furthermore, we derive an analytical expression for evaluation of the expected error probability of a simple detector for the considered system. The accuracy of the proposed analytical expression is verified via particle-based simulation of the Brownian motion.  
\end{abstract}

\section{Introduction} 
Transportation of molecules as a means of conveying information, i.e., molecular communication (MC), is a widely used form of communication in nature. However, only recently has MC attracted the attention of communications researchers in an effort to enable synthetic communication among nano-machines. It is envisioned that networks of communicating nano-machines, i.e., so called nano-networks, can facilitate new revolutionary applications in areas such as biological engineering, healthcare, and environmental engineering \cite{NakanoB}. 

Among the many applications that can potentially benefit from synthetic MC systems, some may require the deployment of \emph{mobile} nano-machines. For instance, in targeted drug delivery and intracellular therapy applications, it is envisioned that mobile nano-machines carry drug molecules, see \cite[Chapter 1]{NakanoB}.  
Knowledge of the channel statistics is crucial in communication design and analysis. However, these statistics change with mobility. Hence, establishing reliable communication between \emph{mobile} nano-machines is more challenging. The design of new modulation, detection, and/or estimation schemes for mobile MC systems requires accurate models for the underlying time-variant channels. 

In the MC literature, the problem of mobile MC has been considered in \cite{108-Luo2016, 109-Qiu2016, 106-Hsu2015, 116-Jamali2016, 105-Guney2012, 107-Kuscu2014, 104-Nakano2016}. In most of these works, only the receiver is mobile (see \cite{108-Luo2016, 106-Hsu2015, 109-Qiu2016, 116-Jamali2016}) and a systematic approach for modeling time-variant channels is not provided. Furthermore, it is assumed that the channel impulse response (CIR) either changes slowly over time, due to the slow movement of the receiver, as in \cite{109-Qiu2016}, or it is fixed for a block of symbol intervals and may change slowly from one block to the next; see \cite{106-Hsu2015, 116-Jamali2016}. In \cite{105-Guney2012} and \cite{107-Kuscu2014}, a \emph{three-dimensional} random walk model is adopted for modeling the mobility of nano-machines, where it is assumed that information is \emph{only} exchanged upon the collision of two nano-machines. In particular, \emph{F\"orster resonance energy transfer} and a \emph{neurospike communication model} have been considered for information exchange between two colliding nano-machines in \cite{105-Guney2012} and \cite{107-Kuscu2014}, respectively. Recently, the authors of \cite{104-Nakano2016} proposed a leader-follower model for target detection applications in two-dimensional mobile MC systems. Langevin equations are used to describe nano-machine mobility. There, it is assumed that the information molecules do not diffuse; the leader nano-machine releases signaling molecules that stick to the release site and form a path that the follower nano-machine follows.

In this letter, we adopt a three-dimensional diffusion model to characterize the movement of both transmitter and receiver nano-machines. Unlike \cite{105-Guney2012} and \cite{107-Kuscu2014}, we assume that nano-machines exchange information via signaling molecules instead of collisions. Furthermore, unlike \cite{104-Nakano2016}, we consider the case where the signaling molecules also \emph{diffuse}, since communication via diffusive signaling molecules is one of the main means of communication among biological entities and therefore also of interest for synthetic MC systems \cite{NakanoB}. We show that by appropriately modifying the diffusion coefficient of the signaling molecules, the CIR of a mobile MC system can be obtained from the CIR of the same system for fixed transmitter and receiver. The proposed solution does not require the stringent assumptions needed in \cite{109-Qiu2016, 106-Hsu2015, 116-Jamali2016} and can accurately account for fast nano-machine movement, even within one symbol interval. Furthermore, we derive an analytical expression for the expected error probability of a simple detector for the considered mobile MC system and verify this expression by particle-based simulation of the Brownian motion and of ligand-receptor binding at the receiver. 

\section{System Model and Preliminaries} 
\label{Sec.SysMod} 
\subsection{System Model}
\begin{figure}[!t] 
	\centering
	\includegraphics[scale = 0.65]{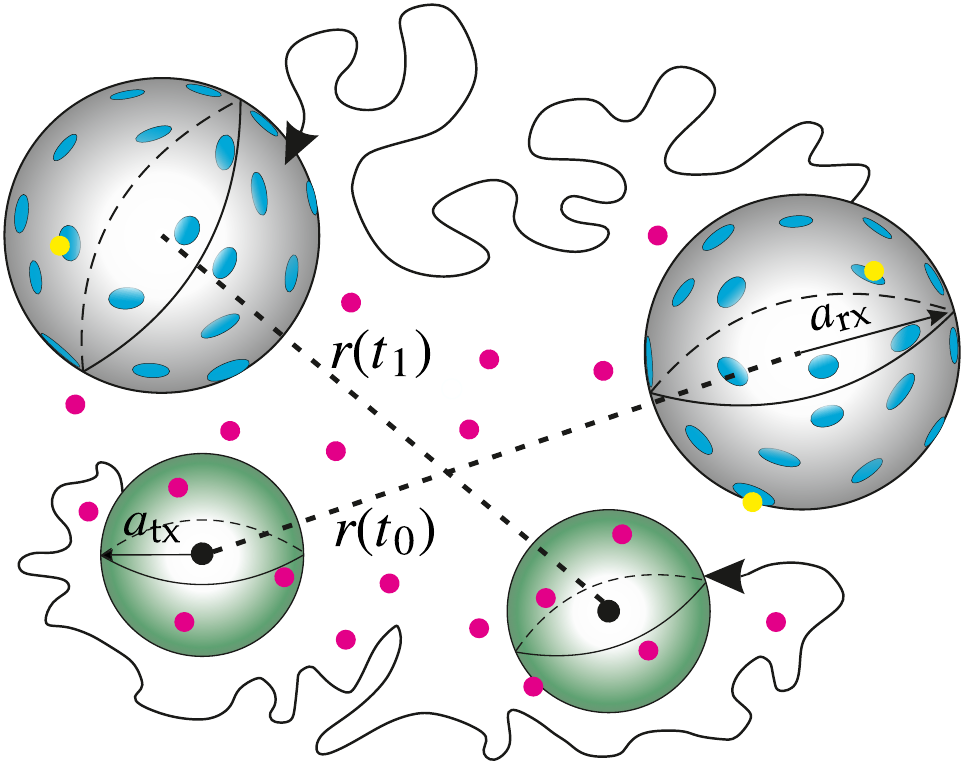}
	\caption{Schematic diagram of the considered system model, where the receiver and the transmitter are shown as a gray and a green sphere, respectively. The unbound and bound information molecules are shown as magenta and yellow dots, respectively. Sample trajectories of the receiver and the transmitter from time $t_0$ to time $t_1$ are shown as solid arrows.} \vspace*{- 5mm} 
	\label{Fig.SystemModel}
\end{figure}

We consider an unbounded three-dimensional fluid environment with constant temperature and viscosity, a spherical transmitter with radius $a_{\text{tx}}$, and a spherical receiver with radius $a_{\text{rx}}$ whose surface is partially covered with $M$ receptor protein molecules, denoted as type $B$ molecules, where we model each receptor as a circular patch with radius $r_s$. Furthermore, we assume that  transmitter and receiver diffuse with constant diffusion coefficients $\diffd{\TX}$ and $\diffd{\RX}$, respectively. We denote the \emph{time-varying} distance between the center of the transmitter and the center of the receiver at time $t$ as $r(t)$, where we assume $r(t_0 = 0) = r_0$; see Fig. \ref{Fig.SystemModel}. Collisions between  transmitter and receiver are assumed to be non-reactive, i.e., upon contact the chemical properties of both remain unaltered, and the transmitter is reflected. As a result, neither the transmitter nor the receiver are \emph{degraded} or \emph{destroyed} while communicating. The transmitter employs type $A$ molecules, that diffuse with constant diffusion coefficient $\diffd{A}$, for conveying information to the receiver. We refer to the $A$ molecules also as information or signaling molecules. We assume that the diffusion process of each $A$ molecule is independent of that of all other $A$ molecules. Moreover, each $A$ molecule can degrade anywhere in the channel via a first order degradation reaction of the form 
\begin{IEEEeqnarray}{C}
	\label{Eq.DegReaction} 
	A \xrightarrow{\kd} \emptyset,  
\end{IEEEeqnarray}
where $\kd$ is the degradation reaction constant in $\text{s}^{-1}$ and $\emptyset$ is a species of molecule that is not recognized by the receiver. Furthermore, we assume that $A$ molecules that reach the vicinity of the receiver may reversibly react with $B$ molecules on the receiver surface and activate them via a second-order reversible reaction as follows 
\begin{IEEEeqnarray}{C}
	\label{Eq.RevReaction} 
	A + B \mathrel{\mathop{\rightleftarrows}^{\kf}_{\kb}} C,  
\end{IEEEeqnarray}
where $\kf$ is the microscopic forward reaction constant in $\text{molecule}^{-1} \text{m}^3 \text{s}^{-1}$, $\kb$ is the microscopic backward reaction constant in $\text{s}^{-1}$, and $C$ denotes an activated receptor. The $A$ molecules cannot penetrate the receiver and are reflected back after unbinding from the receptor (backward reaction) or when they hit a part of the receiver surface that is not covered by receptors. We define the CIR, $\pac{t}$, as the probability that a given $A$ molecule released by the transmitter from $r_0$ at $t_0 = 0$ activates a receptor at time $t$. 
  
Furthermore, we assume that the information that is sent from the transmitter to the receiver is encoded into a binary sequence of length $L$, $\seq = [\bit_1, \bit_2, \cdots, \bit_L ]$. Here, $\bit_j$ is the bit transmitted in the $j$th bit interval with $\pr(\bit_j = 1)= P_1$ and $\pr(\bit_j = 0)= P_0 = 1 - P_1$, where $\pr(\cdot)$ denotes probability. We adopt ON/OFF keying for modulation and a fixed bit interval duration of $T$ seconds. In particular, the transmitter releases a fixed number of $A$ molecules, $N_A$, for transmitting bit ``1'' at the \emph{beginning} of a modulation bit interval and no molecules for transmitting bit ``0''. The signaling molecules are released at the center of the transmitter and leave the transmitter via free diffusion.

\subsection{Preliminaries} 
In this subsection, we present the CIR of the considered system when $\diffd{\TX} = \diffd{\RX} = 0$, i.e., when transmitter and receiver do not move. This case was studied in \cite{ArmanJ2}. Let us assume that the transmitter instantaneously releases $N_A$ $A$ molecules from $r_0$ into the environment at $t_0 = 0$. Since, for $\diffd{\TX} = \diffd{\RX} = 0$, the transmitter and receiver are fixed, $r(t) = r_0$ $\forall$ $t > t_0$. Then, given the above-mentioned assumptions, it has been shown in \cite{ArmanJ2} that the CIR of the system is given by 
\begin{IEEEeqnarray}{rCl} 
	\label{Eq. Channel_Impulse_Response}
 \pac{t} & = & \frac{k_f e^{-k_d t}}{4\pi r_0 a_{\text{rx}} \sqrt{\diffd{A}}} \left\lbrace \frac{\alpha \w\left( \frac{r_0 - a_{\text{rx}}}{\sqrt{ 4 \diffd{A} t}}, \alpha \sqrt{t} \right)}{(\gamma - \alpha)(\alpha - \beta)}  \right. \nonumber \\
 &&  \left. +\> \frac{\beta  \w\left( \frac{r_0 - a_{\text{rx}}}{\sqrt{4 \diffd{A} t}}, \beta \sqrt{t} \right)}{(\beta - \gamma)(\alpha - \beta)}  
  +  \frac{\gamma \w\left( \frac{r_0 - a_{\text{rx}}}{\sqrt{4 \diffd{A} t}}, \gamma \sqrt{t} \right)}{(\beta - \gamma)(\gamma - \alpha)} \right\rbrace, \hspace{3 mm}  	
\end{IEEEeqnarray} 
where $\w( n,m) = \exp(2nm + m^2)\erfc(n+m)$, $\erfc(\cdot)$ denotes the complementary error function, and constants $\alpha$, $\beta$, and $\gamma$ are the solutions to the following system of equations: 
\begin{equation}
	\left\{ \,
	\begin{IEEEeqnarraybox}[][c]{l?s}
	\IEEEstrut
	\alpha + \beta + \gamma = \left( 1 + \frac{\dimv{k}_f}{4 \pi a_{\text{rx}} \diffd{A}} \right) \frac{\sqrt{\diffd{A}}}{a_{\text{rx}}}, \\
	\alpha \gamma + \beta \gamma + \alpha \beta = k_b - k_d, \\
	\alpha \beta \gamma = k_b \frac{\sqrt{\diffd{A}}}{a_{\text{rx}}} - k_d \left( 1 + \frac{\dimv{k}_f}{4 \pi a_{\text{rx}} \diffd{A}} \right) \frac{\sqrt{\diffd{A}}}{a_{\text{rx}}}.
	\IEEEstrut
	\end{IEEEeqnarraybox}
	\right.
	\label{Eq. AlphaBetaGamma}
\end{equation} 
Here, $\dimv{k}_f$ can be evaluated as \cite[Eq. (47)]{ArmanJ2}
\begin{IEEEeqnarray}{C}
	\label{Eq. KfModifiedFinalRelasionship} 
	\dimv{k}_f = \frac{4 \pi \diffd{A} \kf \varphi}{\kf a_{\text{rx}} (1 - \varphi) + 4 \pi \diffd{A}},
\end{IEEEeqnarray} 
where $\varphi$ is a constant that is given by \cite[Eq. (39)]{ArmanJ2}
\begin{IEEEeqnarray}{C} 
	\label{Eq. CorrectionFactor} 
	\hspace{-4mm}\varphi = \frac{M r_s^2 (\kf a_{\text{rx}} + 4 \pi D_A)}{a_{\text{rx}}^2(1 - \lambda)(\pi r_s \kf + 16 \pi D_A) + M r_s^2 (\kf a_{\text{rx}} + 4 \pi D_A)}, 
\end{IEEEeqnarray} 
and $\lambda = M\frac{\pi r_{s}^{2}}{4 \pi a_{\text{rx}}^2}$ is the fraction of the receiver surface covered by receptors. 

Given $\pac{t}$ in \eqref{Eq. Channel_Impulse_Response} and the assumption of independent diffusion of $A$ molecules, the expected received signal at the receiver, $\overline{N}_C(t)$, i.e., the average number of activated receptors at time $t$, is given by 
\begin{IEEEeqnarray}{c} 
	\label{Eq. ExpectedReceSignal} 
	\overline{N}_C(t) = N_A \pac{t}.
\end{IEEEeqnarray} 
\section{Time-Variant Channel Model}
\label{Sec. Time-Variant_CIR} 
In this section, we first present a simple approach for calculation of the CIR of the considered mobile MC system. Then, we derive the probability distribution function (PDF) that describes the evolution of the distance between the transmitter and receiver at the instants when the transmitter releases molecules. Then, given this PDF, we calculate the expected error probability of the considered system for a simple detector.\vspace*{- 4mm}     
\subsection{CIR of a Mobile MC System}
We now assume that $\diffd{\TX} \geq 0$ and $\diffd{\RX} \geq 0$. Furthermore, we still consider only \emph{one impulsive} release of $A$ molecules by the transmitter from $r(t_0) = r_0$ at time $t_0 = 0$. Clearly, after the impulsive release, the diffusion of the transmitter does not influence the CIR. Thus, the knowledge of the position of the transmitter at the molecule release time is sufficient to calculate the CIR of the mobile MC system.

In order to calculate the CIR, one must solve the partial differential equation describing the joint processes of diffusion and degradation of signaling molecules in the channel under \emph{time-variant} boundary conditions describing the reversible reaction of the $A$ molecules with the receptor $B$ molecules on the surface of a \emph{diffusive} receiver. However, the mobility of the receiver, and the resulting time variation of the boundary conditions required for describing \eqref{Eq.RevReaction}, makes the problem of finding the CIR very difficult and analytically intractable. In order to overcome this difficulty, we adopt the concept of relative diffusion of two particles from \cite{H.KimJ1}. In particular, in \cite{H.KimJ1}, it is shown that the reversible reaction of two molecule species, namely $q_1$ and $q_2$, with corresponding diffusion coefficients $\diffd{q_1}$ and $\diffd{q_2}$, can be accurately described by assuming that either of the two particles is static and the other one diffuses with an effective diffusion coefficient that is the summation of the diffusion coefficients of the individual molecules, i.e., $\diffd{\text{eff}} = \diffd{q_1} + \diffd{q_2}$.

For the problem at hand, however, the receptor $B$ molecules are mounted on the surface of the receiver and, as a result, undergo Brownian motion with diffusion coefficient $\diffd{\RX}$. Thus, we introduce an effective diffusion coefficient that describes the relative motion of a given $A$ molecule with respect to the motion of a given $B$ receptor molecule as 
\begin{IEEEeqnarray}{C}
	\label{Eq. DiffCoeffEff} 
	\diffd{\text{eff},1} = \diffd{A} + \diffd{\RX}.
\end{IEEEeqnarray}          

Finally, the CIR and the expected received signal for the mobile transmitter and receiver scenario are given by \eqref{Eq. Channel_Impulse_Response} and \eqref{Eq. ExpectedReceSignal}, respectively, after substituting $\diffd{A}$ with $\diffd{\text{eff},1}$ given in \eqref{Eq. DiffCoeffEff}.
\vspace*{- 3mm}
\begin{remark}
The approach proposed here for evaluation of the CIR of a mobile MC system is general and can be also applied to other receiver models, e.g., passive (i.e., non-reactive) and perfectly-absorbing models \cite{NoelPro2}.
\end{remark}
\vspace*{- 4mm} 
\subsection{Relative Diffusion of Transmitter and Receiver Nodes}
\label{Sec. RelDiffTxRx} 
As stated in Section \ref{Sec.SysMod}, we treat transmitter and receiver as two non-reactive particles. In other words, we assume that the transmitter bounces off the receiver upon collision. Then, we are interested in finding $\pr(r(t) = r | r_0)$, i.e., the probability that at time $t$, $r(t)$ is equal to a sample distance $r \geq a_{\text{rx}} + a_{\text{tx}}$, given that $r(t_0=0)=r_0$. For two non-reactive particles, $\pr(r(t) = r | r_0)$ can be calculated from \cite[Eq. (17)]{H.KimJ1} and be written as
\begin{IEEEeqnarray}{rCl}
	\label{Eq. RelMovTxRx}
	\pr\left( r(t) = r | r_0 \right) & = & \frac{\exp\left(\frac{-\left( r - r_0 \right)^2}{4 t \diffd{\text{eff},2}}\right) + \exp\left( \frac{-\left( r + r_0 - 2(a_{\text{rx}}+ a_{\text{tx}}) \right)^2}{4 t \diffd{\text{eff},2}} \right)}{ \frac{r_0}{r}\sqrt{ 4 \pi t \diffd{\text{eff},2}} }  \nonumber \\  
	 && -\>  \frac{r}{r_0}\w\left( \frac{r + r_0 - 2(a_{\text{rx}}+ a_{\text{tx}})}{\sqrt{ 4 t \diffd{\text{eff},2}}}, \sqrt{t} \right).	 
\end{IEEEeqnarray}     
Here, similar to \eqref{Eq. DiffCoeffEff}, we employ an effective diffusion coefficient, $\diffd{\text{eff},2}$, to characterize the relative motion of the transmitter and receiver, where $\diffd{\text{eff},2}$ is given by 
\begin{IEEEeqnarray}{C}
	\label{Eq. DiffCoeffEff2} 
	\diffd{\text{eff},2} = \diffd{\TX} + \diffd{\RX}.
\end{IEEEeqnarray}  
It can be shown that $\int_{a_{\text{rx}} + a_{\text{tx}}}^{+ \infty} \pr\left( r(t) = r | r_0 \right) \dif r = 1$. 
  
Now, let us define vector $\rseq = [r[1], r[2], \cdots, r[L-1]]$ whose $\kappa$th element is defined as $r[\kappa] = r(t)|_{t = \kappa T}$. Then, the $(L-1)$-dimensional joint PDF $f_{\rseq}(\vec{r})$, where $\vec{r} = [r_0, r_1, \cdots, r_{L-1}]$ is one sample realization of $\rseq$, is given by 
\begin{IEEEeqnarray}{rCl}
	\label{Eq. JointPDF} 
	f_{\rseq}(\vec{r}) & = & \prod_{i = 1}^{L-1} \pr\big(r[i] = r_{i}| r_{i-1}\big), 
\end{IEEEeqnarray} 
where we exploited a) $\pr(r(t_0=0) = r_0 | r_0) = 1$, and b) $\pr(r[i]=r_i| r_0, r_1, \cdots, r_{i-1}) = \pr(r[i]=r_i|r_{i-1})$. \vspace*{- 3mm}      
\subsection{Expected Error Probability of Mobile MC System}
\label{Sec. ExpBER} 
In this subsection, we derive an analytical expression for the expected error probability of mobile MC systems employing a simple detector. To this end, we first provide the expected error probability of the mobile MC system for \emph{any} detector at the receiver. Let us assume that $\seq$ and $\vec{r}$ are given. We define $\Pe(\bit_j | \seq, \vec{r}) = \pr(\hat{\bit}_j \neq \bit_j | \seq, \vec{r})$ as the conditional error probability of the $j$th bit, where $\hat{\bit}_j$ denotes the $j$th detected bit. Then, given $\Pe(\bit_j | \seq, \vec{r})$ and $f_{\rseq}(\vec{r})$, the \emph{expected} error probability of the $j$th bit, $\mathrm{\overline{P}_e}(\bit_j)$, can be calculated as  
\begin{IEEEeqnarray}{rCl}
	\label{Eq. ExpErrorProb}
	\mathrm{\overline{P}_e}(\bit_j) = \idotsint\limits_{\vec{r} \in \mathcal{R}} \sum_{\seq \in \mathcal{B}} f_{\rseq}(\vec{r}) \pr(\seq) \Pe(\bit_j | \seq, \vec{r}) \dif r_1 \dif r_2 \cdots \dif r_{L-1}, \nonumber \\* 
\end{IEEEeqnarray}
where $\mathcal{R}$ and $\mathcal{B}$ are sets containing all possible realizations of $\vec{r}$ and $\seq$, respectively, and $\pr(\seq)$ is the likelihood of the occurrence of $\seq$. 

\begin{figure*}[!tbp]
\begin{minipage}{\textwidth}
  \begin{minipage}[b]{0.24\textwidth}
  	\begin{scriptsize}
  	\renewcommand{\arraystretch}{1.15} 
    \centering 
    \begin{tabular}{|c||c|}\hline 
      Param. & Value \\ \hline \hline
        $N_A$ & $5000$ \\ \hline
        $D_A$ & $0.5 \times 10^{-9}$ $\text{m}^2/{\text{s}}$ \\ \hline
        $D_{\text{RX}}$ & $0.5 \times 10^{-12}$ $\text{m}^2/{\text{s}}$ \\ \hline
        $r_0$ & $1$\, $\mu$m \\ \hline
        $a_{\text{rx}}$   & $0.5$\, $\mu$m \\ \hline
        $\kf$ &  $12.5 \times 10^{-15}$ $\frac{\text{m}^3}{\text{molecule $\cdot$ s}}$ 												\\ \hline
        $\kb$ & $2 \times 10^{5}$ $\text{s}^{-1}$ \\ \hline
        $\kd$ & $0.2 \times 10^{5}$ $\text{s}^{-1}$ \\ \hline
        $M$   & $1000$ \\ \hline
        $r_s$ & $13.95$ nm \\ \hline
        $T$ & $0.3$ ms \\ \hline
        $t_s$ & $0.06$ ms \\ \hline
        $L$   & $10$ \\ \hline
      \end{tabular}
      \captionof{table}{Simulation \hspace{-1mm} Parameters.} \vspace*{- 5mm} 
      \label{Tabel}
      \end{scriptsize}
    \end{minipage}
    \hfill 
    \begin{minipage}[b]{0.37\textwidth}
    	\centering
    	\includegraphics[scale = 0.25]{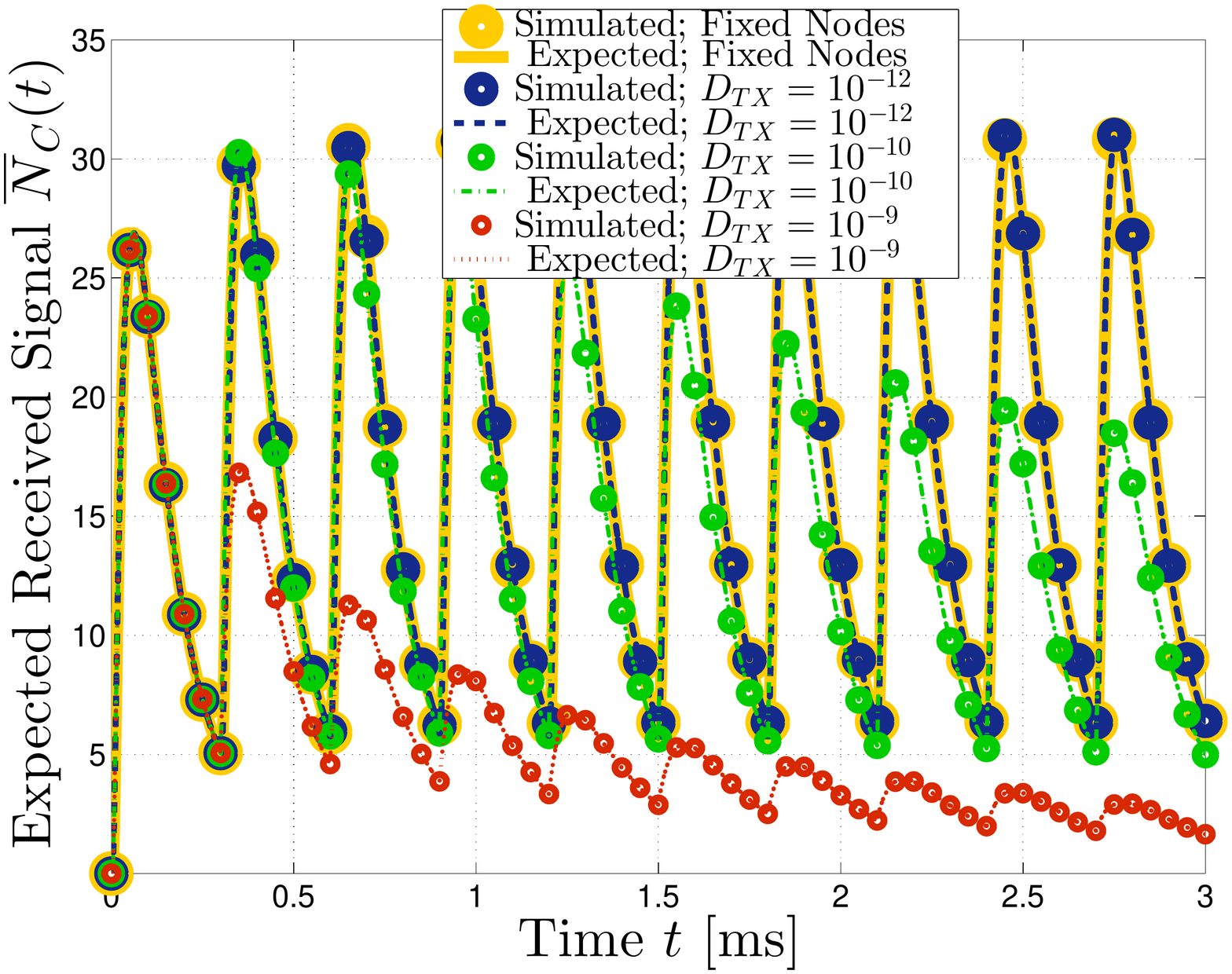}
    	\caption{$\overline{N}_C(t)$ as a function of time $t$.}\vspace*{- 5mm}  
    	\label{Fig. Analysis1}
  	\end{minipage}
  	\hfill
  	\begin{minipage}[b]{0.37\textwidth}
    	\centering
    	\includegraphics[scale = 0.25]{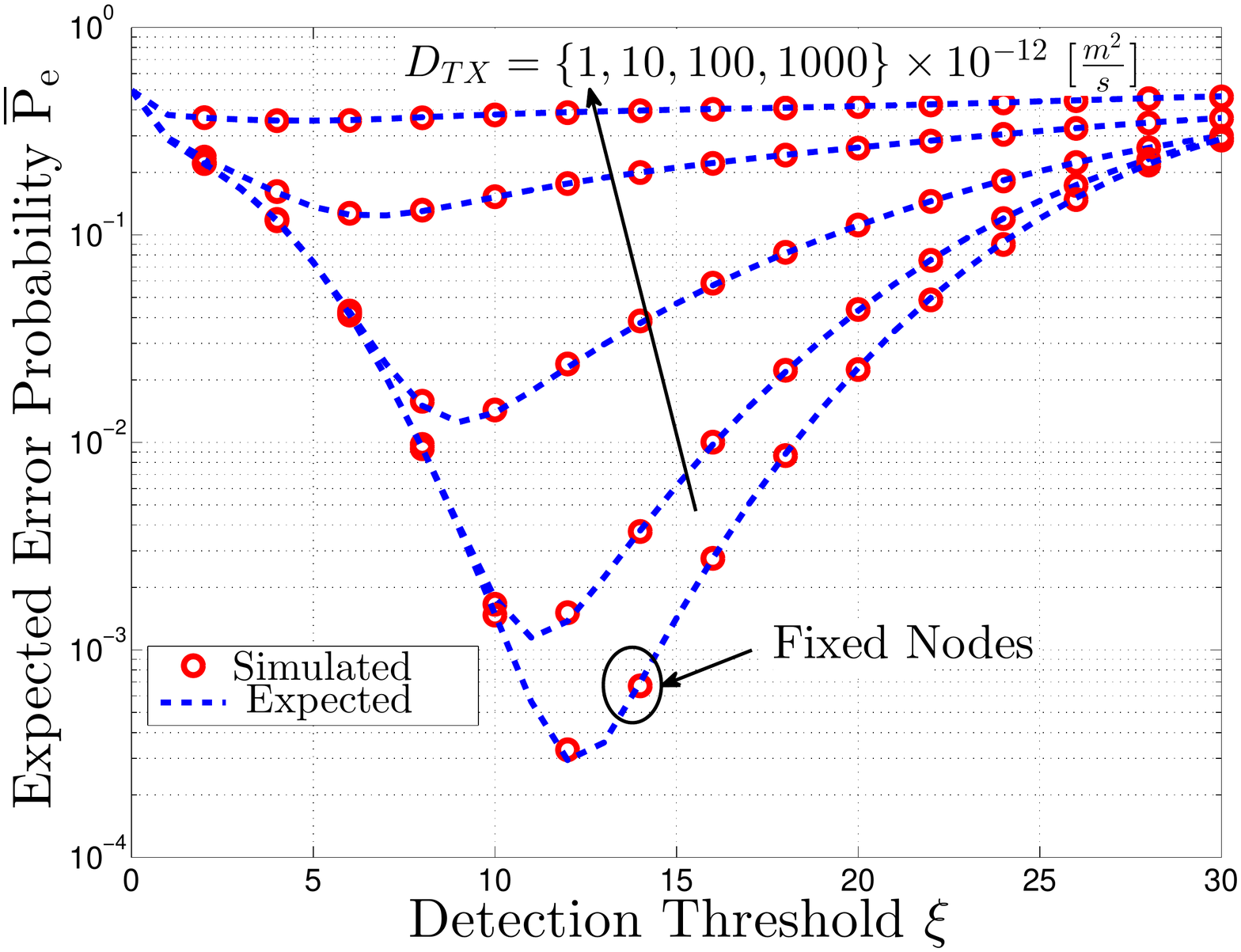}
    	\caption{$\mathrm{\overline{P}_e}$ as a function of detection threshold, $\xi$.}\vspace*{- 5mm} 
    	\label{Fig. Analysis2}
  	\end{minipage}
  	\hfill
  \end{minipage}
  \label{FigurelanaT}
\end{figure*}       

In the remainder of this section, we adopt a \emph{single-sample} detector and show how $\mathrm{P_e}(\bit_j| \seq, \vec{r})$ can be evaluated in this case. In particular, in each symbol interval, the receiver counts the number of activated receptors at a fixed sampling time after the beginning of the symbol interval, $t_s$, and compares the number of $C$ molecules $N_C(t_{j,s})$ with a fixed threshold value $\xi$ to make a decision as follows: 
\begin{equation}
	\label{Eq.Reception} 
	\hat{\bit}_j = \begin{cases} 
	1 &\mbox{if } N_C(t_{j,s}) \geq \xi, \\
	0 &\mbox{if } N_C(t_{j,s}) < \xi, 
			\end{cases}
\end{equation}
where $t_{j,s} = (j-1)T + t_s$. From the perspective of the receiver, at any given time $t$ after an impulsive release of $A$ molecules, any receptor $B$ molecule is either activated or not activated. Thus, the receiver observations of an impulsive release can be modeled as a Binomial random variable and accurately approximated by a Poisson distribution when the number of trials ($N_A$) is large and the probability of success is small ($P_{AC}(\cdot) \ll 1$). Given $\seq$ and $\vec{r}$, the mean of the received signal at the reactive receiver, denoted by $\overline{N}_C(t_{j,s})$, can be expressed as 
\begin{IEEEeqnarray}{C}
	\label{Eq. PoissonMean} 
	\overline{N}_C(t_{j,s}) = N_A \sum_{i=1}^{j}\bit_i P_{AC}((j-i)T + t_s|r_{i-1}),
\end{IEEEeqnarray} 
where we exploited the fact that a sum of independent Poisson random variables is also a Poisson random variable, given that the bits in sequence $\seq$ are independent of each other. Then, given decision rule \eqref{Eq.Reception}, $\mathrm{P_e}(\bit_j| \seq, \vec{r})$ can be written as 
\begin{IEEEeqnarray}{C}
	\label{Eq. ExpErrorProbDetector} 
	\mathrm{P_e}(\bit_j| \seq, \vec{r}) = \begin{cases} 
	\pr(N_C(t_{j,s}) < \xi) &\mbox{if } \bit_j = 1, \\
	\pr(N_C(t_{j,s}) \geq \xi) &\mbox{if } \bit_j = 0, 
			\end{cases}
\end{IEEEeqnarray} 
where $\pr(N_C(t_{j,s}) < \xi)$ can be calculated from the cumulative distribution function of a Poisson distribution as 
\begin{IEEEeqnarray}{rCl}
	\label{Eq. PoissCDF} 
\hspace{-3 mm}	\pr(N_C(t_{j,s}) < \xi) = \exp\left(-\overline{N}_C(t_{j,s})\right) \sum_{\omega = 0}^{\xi - 1} \frac{\left(\overline{N}_C(t_{j,s})\right)^{\omega}}{\omega !},
\end{IEEEeqnarray}
and $\pr(N_C(t_{j,s}) \geq \xi) = 1 - \pr(N_C(t_{j,s}) < \xi)$. Given $\mathrm{P_e}(\bit_j| \seq, \vec{r})$ in \eqref{Eq. ExpErrorProbDetector}, $\mathrm{\overline{P}_e}(\bit_j)$ can be calculated via \eqref{Eq. ExpErrorProb}. Finally, the expected error probability of the mobile MC system can be evaluated by averaging over all bit intervals, i.e., $\mathrm{\overline{P}_e} = \frac{1}{L} \sum_{j=1}^{L} \mathrm{\overline{P}_e}(\bit_j)$, where we used Monte-Carlo simulation for evaluation of the multi-dimensional integral in \eqref{Eq. ExpErrorProb}.  
\section{Simulation Results}
\label{Sec. SimulationResults} 
In this section, we present simulation and numerical results. 
For simulation, we extended the Brownian motion particle-based framework proposed in \cite{ArmanJ2}. In particular, in our extension, we added a three-dimensional random walk model for the movement of the transmitter and receiver nodes. We treated the transmitter and the reactive receiver as two hard spheres, i.e., we assumed that they cannot occupy the same space at the same time, and upon collision the transmitter is reflected. 

For all simulation results, we chose the set of simulation parameters provided in Table \ref{Tabel}. Furthermore, we considered a water environment at $25\, \mathrm{{}^{\circ}C}$ and used the Stokes--Einstein equation \cite[Eq.~(5.7)]{NakanoB} for calculation of $D_A$, $\diffd{\RX}$, and $\diffd{\TX}$. The only parameters that were varied are $\xi$ and $\diffd{\TX}=\{1, 10, 100, 1000\}\times 10^{-12}$ $\frac{\text{m}^2}{\text{s}}$ (corresponding to $a_{\text{tx}} = 0.24357 \times \{10^{-6}, 10^{-7}, 10^{-8}, 10^{-9}\}$ m). All simulation results were averaged over $15 \times 10^{3}$ independent realizations of the environment. In the following, we refer to the scheme where $\diffd{\TX} = \diffd{\RX} = 0$ as ``fixed nodes''.

Fig.~\ref{Fig. Analysis1} shows the expected received signal, $\overline{N}_C(t)$, as a function of time $t$ for transmission of $L=10$ consecutive ``1''s. For clarity of exposition, we do not show the results for $\diffd{\TX} = 10^{-11}$. First, we observe that in each bit interval, as expected due to the impulsive release of $N_A$ $A$ molecules, $\overline{N}_C(t)$ first increases with time and then decreases. Furthermore, as time $t$ increases, the signal received in each bit interval starts to decrease when $\diffd{\TX} > 0$ and/or $\diffd{\RX} > 0$. This is mainly due to the fact that when $\diffd{\RX}>0$ and/or $\diffd{\TX}>0$, the transmitter and the receiver ultimately diverge and, as a result, $\overline{N}_C(t \to \infty) \to 0$. For larger $\diffd{\text{eff,2}} = \diffd{\TX} + \diffd{\RX}$, due to the faster movement of the transmitter and/or receiver, $r(t)$ increases faster and $\overline{N}_C(t)$ tends to zero sooner.

Fig.~\ref{Fig. Analysis2} presents the expected error probability, $\mathrm{\overline{P}_e}$, as a function of detection threshold $\xi$ for $L=10$ when $P_1 = P_0 =0.5$. Fig.~\ref{Fig. Analysis2} shows that for all considered cases, the expected error probability first decreases with increasing $\xi$ and then increases. This is because by increasing $\xi$, $\pr(\hat{\bit}_j \neq \bit_j|\bit_j = 0)$ decreases and $\pr(\hat{\bit}_j \neq \bit_j|\bit_j = 1)$ increases. Furthermore, for increasing $\diffd{\text{eff,2}}$, the overall performance of the system deteriorates. This is because for larger values of $\diffd{\TX}$, and consequently larger values of $\diffd{\text{eff,2}}$, as observed in Fig.~\ref{Fig. Analysis1},  $\overline{N}_C(t)$ tends to zero faster. We also observe that even though the $\overline{N}_C(t)$ for the ``fixed nodes'' and $\diffd{\TX} = 10^{-12}$ are approximately the same (see Fig.~\ref{Fig. Analysis1}), their BERs are different as small differences in $\overline{N}_C(t)$ have a large impact on the BER.

Finally, for both Figs.~\ref{Fig. Analysis1} and \ref{Fig. Analysis2}, we note the excellent match between simulation and analytical results.   
\vspace{-2 mm}
\bibliography{IEEEabrv,Library}
\end{document}